\documentclass[aps,prd,onecolumn,10pt,groupedaddress,nofootinbib,showpacs]{revtex4}
\usepackage[utf8x]{inputenc}
\usepackage{amsmath,amssymb}
\usepackage[dvips]{graphicx}
\usepackage{pstricks}
\usepackage{comment}
\makeindex

\begin{document}


\author{Tamás Temesv\'ari}\email{temtam@helios.elte.hu}
\affiliation{Theoretical Physics Department, Institute of Physics,
E\"otv\"os Loránd University, P\'azm\'any P\'eter s\'et\'any 1/A,
H-1117 Budapest, Hungary
}
\author{Imre Kondor}\affiliation{Complexity Science Hub, Vienna, Austria and
London Mathematical Laboratory, London, UK}

\title{Field theory for the Almeida-Thouless transition}
\begin{abstract}
 This work contains a review of the theoretical understanding for the Almeida-Thouless
 transition. 
 The effective field theoretic model for this transition 
 to the replica symmetry broken phase
 is extended to low temperatures, making it possible to compute perturbative corrections to the mean field phase boundary
 even close to zero temperature. Nonperturbative behavior emerges both around the 
 zero-external-field transition and at low temperature. Renormalization group results are
 also discussed.
\end{abstract}

\maketitle

\section{Introduction}
 
 Mean field (MF) theory is exact for the Ising spin glass on the fully connected lattice,
i.e.\  the SK model, and its simplest solution has a transition from the paramagnet to the
replica symmetric (RS) SG state in zero external field \cite{SK}. This RS phase, however, was soon proven to be
unstable for zero as well as for any nonzero magnetic field
whenever the temperature is low enough \cite{AT}. This instability is now understood to indicate the onset of replica symmetry breaking (RSB) in both cases. Yet the nature of the instability differs in one from the other:
\begin{itemize}
 \item $H=0\,$: The high-temperature paramagnetic phase has a unique
 degenerate mass $m$ [with multiplicity $n(n-1)/2$ in the replicated theory, with $n$ being the replica number]. The MF transition at the SG critical point $T_c^{\text{mf}}$ has the character of a paramagnet to RSB SG transition, instability of the paramagnet is signaled by $m\to 0$.
 \item $H>0\,$:  The high-$T$ phase has three different masses: replicon $m_R$,  anomalous $m_A$, and longitudinal $m_L$. (For $n=0$, the latter two are degenerate.) Upon lowering $T$, $m_R$ vanishes at the Almeida-Thouless (AT) instability, whereas the other two modes remain noncritical. This kind of transition
 is now considered to be the \emph{true} SG transition, physically resulting in the SG susceptibility to diverge, whereas the zero-field case is multicritical.
\end{itemize}

\begin{figure}
\centerline{\includegraphics[scale=0.4]{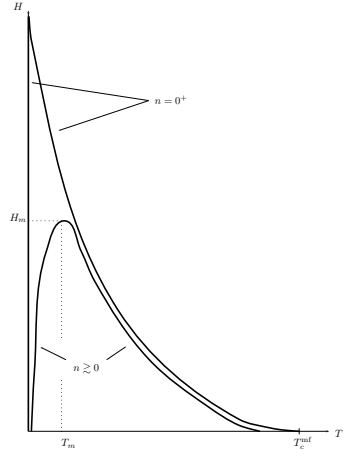}}
\caption{MF phase diagram of the Ising spin glass for $n=0^+$ and $n\gtrsim 0$. The maximum asymptotically scales as
$T_m\sim n\,[\ln(n^{-2})]^{1/2}$ and $H_m\sim [\ln(n^{-2})]^{1/2}$. The low-$T$ AT line terminates at $T_{\text{min}}\sim n\,[\ln (n^{-2})]^{-1/2}$ (not marked in the figure) for $H=0$,  signalling the RS reentrance.}
\label{fig1}
\end{figure}

An AT transition can take place even when $H=0$, but only for $n\gtrsim 0$ \cite{Kondor}. By extending the finite $n$ calculation to $H>0$, one can contrast the phase diagram with that for $n=0$ (Fig.~\ref{fig1}). We note the following:
\begin{description}
 \item[(i)] The finite $n$ phase boundary has a maximum, and hence the RS phase reenters at low $T$.
 \item [(ii)]The high-$T$ endpoint of the finite $n$ AT line
 is separated from $T_c^{\text{mf}}$ by a \emph{stable} RS SG phase.
 \item[(iii)] The $T\to 0$ and $n\to 0$ limit is strongly singular: the  $H$ axis for $n=0$ is simultaneously the innermost part of the RSB phase, and the $n\to 0$ limit of the low-$T$ wing of the AT line.
\end{description}

\section{RS field theory for the AT transition}
\label{the_model}

Going beyond MF theory in $d$ dimensions is commonly
done by building an effective field theory which is suited
to the calculation of perturbative corrections, and may be considered as an initial condition for iterating the renormalization group flows. For this purpose, it is usual to apply the Gaussian integral representation of the replicated and averaged partition function \cite{BrMo79,BrRo,rscikk}:
\[
 \overline{Z^n}\sim\int[d\phi]\:e^{-\mathcal L(\phi)}\:,\qquad
 \mathcal L(\phi)=\mathcal{L}^{(G)}(\phi)+\mathcal{L}^{(I)}(\phi)
\]
with
\begin{equation}\label{LG}
\mathcal{L}^{(G)}=\frac{1}{2}\sum_{\mathbf p}\bigg[
\Big(\frac{1}{2} (pa\rho)^2+m_1\Big)\sum_{\alpha\beta}
\phi^{\alpha\beta}_{\mathbf p}\phi^{\alpha\beta}_{-\mathbf p}
+m_2\sum_{\alpha\beta\gamma}
\phi^{\alpha\gamma}_{\mathbf p}\phi^{\beta\gamma}_{-\mathbf p}\\+
m_3\sum_{\alpha\beta\gamma\delta}
\phi^{\alpha\beta}
_{\mathbf p}\phi^{\gamma\delta}_{-\mathbf p}
\bigg],
\end{equation}
and
\begin{equation}\label{LI}
 \mathcal{L}^{(I)}=-\frac{1}{3!\,\sqrt{N}}
\sideset{}{'}\sum_{\mathbf {p_1p_2p_3}}\:
\sum_{i=1}^8 w_i\,I^{(3)}_i(\phi)
-\frac{1}{4!\,N}\:
\sideset{}{'}\sum_{\mathbf {p_1p_2p_3p_4}}
\sum_{i=1}^{23} u_i\,I^{(4)}_i(\phi)-\dots
\end{equation}
 where the fluctuating fields obey $\phi^{\alpha\beta}_{\mathbf p}=\phi^{\beta\alpha}_{\mathbf p}$ with $\phi_{\mathbf p}^{\alpha\alpha}=0$, the number of the lattice sites $N\to\infty$ in the thermodynamic limit, while $\rho a$
 is the interaction range. One can define the effective coordination number as $z\equiv \rho^d$. (Momentum conservation is understood in the primed sums.) The cubic and quartic RS invariants\footnote{$I^{(k)}_j$ are deemed RS invariant because they are unaffected by the global transformation 
 ${\phi'}^{\alpha\beta}
_{\mathbf p}=\phi^{P_\alpha P_\beta}
_{\mathbf p}$, where $P$ is any permutation of the $n$ replicas.}
 in the interaction Lagrangian $\mathcal{L}^{(I)}$
 have been exhibited in Refs.\ \cite{rscikk,nucl}, some examples are displayed below:
 \begin{gather*}
 I^{(3)}_1(\phi)=\sum_{\alpha\beta\gamma}\phi^{\alpha\beta}
_{\mathbf p_1}\phi^{\beta\gamma}_{\mathbf p_2}
\phi^{\gamma\alpha}_{\mathbf p_3}\:,\quad
I^{(3)}_2(\phi)=\sum_{\alpha\beta}\phi^{\alpha\beta}
_{\mathbf p_1}\phi^{\alpha\beta}
_{\mathbf p_2}\phi^{\alpha\beta}_{\mathbf p_3}\:,\quad
I^{(3)}_3(\phi)=\sum_{\alpha\beta\gamma}\phi^{\alpha\beta}
_{\mathbf p_1}\phi^{\alpha\beta}
_{\mathbf p_2}\phi^{\alpha\gamma}_{\mathbf p_3}\:,\\
\intertext{and}\\
I^{(4)}_1=\sum_{\alpha\beta\gamma\delta}\phi^{\alpha\beta}
_{\mathbf p_1}\phi^{\beta\gamma}_{\mathbf p_2}
\phi^{\gamma\delta}_{\mathbf p_3}\phi^{\delta\alpha}_{\mathbf p_4}
\:,\quad
I^{(4)}_2=\sum_{\alpha\beta}\phi^{\alpha\beta}_{\mathbf p_1}\phi^{\alpha\beta}
_{\mathbf p_2}\phi^{\alpha\beta}_{\mathbf p_3}\phi^{\alpha\beta}_{\mathbf p_4}
\:,\quad
I^{(4)}_5=\sum_{\alpha\beta\gamma}\phi^{\alpha\beta}
_{\mathbf p_1}\phi^{\alpha\beta}_{\mathbf p_2}
\phi^{\alpha\gamma}_{\mathbf p_3}\phi^{\beta\gamma}_{\mathbf p_4}\:.
\end{gather*}
The stationary condition, which requires the linear term in the interaction part to vanish, gives that masses and couplings
depend on temperature $T$, magnetic field $H$ and replica number $n$, thus providing an effective field theory beyond the close vicinity of the zero-field multicritical point. One can therefore also study the $T\to 0$ regime. Exact relations further relate some couplings, the most important being $w_3=-3w_2=-2w_5$.

A hierarchy of the masses and couplings emerges close to $T_c^{\text{mf}}$.
 For the paramagnet, $m_1$, $w_1$, $u_1$,
$u_2$, and $u_3$ are the only nonzero bare parameters (up to quartic order). In the crossover region, the Lagrangian can be written as $\mathcal L=\mathcal L_{\text{para}}+\delta\mathcal L$
with $\delta\mathcal L$ having additionally the parameters $m_2$,
$w_2$, $w_3$, $w_5$, $u_5$ etc., all of which are proportional to the
reduced temperature $\tau=(T_c^{\text{mf}}-T)/T_c^{\text{mf}}$.
It is clear that
the above representation of the system by the RS invariants with
unrestricted replica summations and couplings belonging to them
is well suited to the system
close to the zero-field multicritical point where the paramagnet becomes unstable.

Close to $T=0$, however, a new system of couplings must be chosen. We then decompose the fluctuating field as
\[
 \phi^{\alpha\beta}_{\mathbf p}=(\phi^R_{\mathbf p})^{\alpha\beta}
 +(\phi^A_{\mathbf p})^{\alpha\beta}+(\phi^L_{\mathbf p})^{\alpha\beta}
\]
where
\begin{itemize}
 \item the replicon (R) field has the property $\sum_\beta(\phi^R_{\mathbf p})^{\alpha\beta}=0$ for any $\alpha$, so
 the number of independent components is $n(n-3)/2$;
 \item the anomalous (A) field can be built up from $n-1$
 one-replica fields $(\phi^A_{\mathbf p})^{\alpha}$ with the property $\sum_\alpha(\phi^A_{\mathbf p})^{\alpha}=0$ as
 $(\phi^A_{\mathbf p})^{\alpha\beta}=\frac{1}{2}\big[(\phi^A_{\mathbf p})^{\alpha}+
 (\phi^A_{\mathbf p})^{\beta}\big]$, $\alpha\not=\beta$;
 \item the single component longitudinal (L) field is constant: $(\phi^L_{\mathbf p})^{\alpha\beta}=(\phi^L_{\mathbf p})$, $\alpha\not=\beta$.
\end{itemize}
$\mathcal{L}^{(G)}$ is diagonal in this new representation, whereas the cubic part of $\mathcal{L}^{(I)}$ takes the form
\begin{multline}\label{cubic_with_g}
 -\frac{1}{3!\,\sqrt{N}}
\sideset{}{'}\sum_{\mathbf {p_1p_2p_3}}\bigg\{
g_1\cdot\sum_{\alpha\beta\gamma}(\phi^R_{\mathbf p_1})^{\alpha\beta}(\phi^R_{\mathbf p_2})^{\beta\gamma}(\phi^R_{\mathbf p_3})^{\gamma\alpha}+\frac{1}{2}g_2\cdot \sum_{\alpha\beta}
(\phi^R_{\mathbf p_1})^{\alpha\beta}
(\phi^R_{\mathbf p_2})^{\alpha\beta}(\phi^R_{\mathbf p_3})^{\alpha\beta}\\[8pt]
+3g_3\cdot \sum_{\alpha\beta}(\phi^R_{\mathbf p_1})^{\alpha\beta}(\phi^R_{\mathbf p_2})^{\alpha\beta}(\phi^A_{\mathbf p_3})^{\alpha}
+3g_4\cdot \sum_{\alpha\beta}(\phi^R_{\mathbf p_1})^{\alpha\beta}(\phi^R_{\mathbf p_2})^{\alpha\beta}(\phi^L_{\mathbf p_3})+3g_5\cdot \sum_{\alpha\beta}(\phi^R_{\mathbf p_1})^{\alpha\beta}(\phi^A_{\mathbf p_2})^{\alpha}(\phi^A_{\mathbf p_3})^{\beta}\\[8pt]
+g_6\cdot \sum_{\alpha}(\phi^A_{\mathbf p_1})^{\alpha}
(\phi^A_{\mathbf p_2})^{\alpha}(\phi^A_{\mathbf p_3})^{\alpha}
+3g_7\cdot \sum_{\alpha}(\phi^A_{\mathbf p_1})^{\alpha}
(\phi^A_{\mathbf p_2})^{\alpha}(\phi^L_{\mathbf p_3})+
g_8\cdot(\phi^L_{\mathbf p_1})(\phi^L_{\mathbf p_2})(\phi^L_{\mathbf p_3})
\bigg\}\:.
\end{multline}
See \cite{rscikk} for the relation between the two sets of
couplings, the $g_i$'s and the $w_i$'s.

To calculate corrections to MF theory near $T=0$,
we must know how the bare parameters behave in its vicinity along the MF (or tree-approximation) AT line. It is then convenient to study the $n=0$ and $n\gtrsim 0$ cases separately (see Fig.~\ref{fig1}).
\begin{itemize}
 \item$n=0$:
 The two fully-replicon cubic vertices (i.e.\ with all the three legs being
 R) diverge as
 \begin{equation}\label{replicon_vertices}
  g_1=g_2\sim T^{-1}\:,
  \end{equation}
 whereas the others vanish like $\sim T\ln T$. Surprisingly, the longitudinal mass $m_L=m_A$ does \emph{not}
 become infinitely large in this limit, instead $\lim_{T\to 0}m_L=O(1)$. Interestingly, it is not monotonic along the AT line, but has a maximum at some intermediate temperature.
\item$n\gtrsim0$:
In the low temperature regime, where
the $n\gtrsim0$ line deviates from the $n=0$ one,
the two replicon vertices behave again as in Eq.~\eqref{replicon_vertices}. The other six vertices behave at most as $g_i\sim T^{-1}\cdot n^2$. The AT line, however, reaches the temperature axis ($H=0$) at $T_{\text{min}}\sim n\,[\ln (n^{-2})]^{-1/2}$, and the $n\to 0$ limit finally makes these vertices vanish.
\end{itemize}

\section{Perturbative correction to the mean field AT line}
\label{perturbative_part}

Because perturbative considerations are somewhat modified at $d=6$,
our study in this subsection is restricted to $d>6$.

\begin{description}

\item[(i)] 
{\bfseries The high-temperature endpoint of the AT line for
$H=0$:}

When both the replica number $n$ and the magnetic field $H$
are zero, the replicon mass is negative, $m_R=-\frac{4}{3}\,\tau^2$, thus yielding an ill-defined replicon propagator. We must therefore resort to regularization by $n$ or $H$:

\begin{itemize}
 \item $n\gtrsim 0$ and $H=0$.
 The RS phase is stable between $\tau_c$ and $\tau_{\text{AT}}$,
 and
it can also be proved that $\tau_c$ is at the same time the temperature (at one-loop level) where the paramagnet becomes unstable and the RS order parameter changes sign from negative to positive value.

As for $\tau_{\text{AT}}$, applying \emph{conventional} perturbative method with
$1/z\ll n\ll 0$ at one-loop order and 
contemplating the higher order corrections suggests the 
form
\[
 \tau_{\text{AT}}=n\cdot f_1(1/nz)+n^2\cdot f_2(1/nz)+\dots\:,
\]
and by fixing $z$ while $n\to 0$, the high-argument limit of the $f$ functions will yield the $1/z$ expansion of
$\tau_{\text{AT}}(H=0)$. 

\item $n=0$ and $H^2/(kT_c^{\text{mf}})^2\gtrsim 0$. In this case,
one can compute the AT temperature for a given, small magnetic field perturbatively:
\[
 \tau_{\text{AT}}=\tau_0+O(1/z)\,,
\qquad\text{with}\qquad
 \tau_0\equiv \left[\frac{3}{4}\,\frac{H^2}{(kT_c^{\text{mf}})^2}\right]^{1/3}\:.
\]
The loop expansion is generated for a given, albeit small, magnetic field with $1/z\ll \tau_0\ll 1$. One can expect that a resummation of the whole series provides
\[
 \tau_{\text{AT}}=\tau_0\cdot \bar f(1/\tau_0 z)+\text{correction terms}\:,
\]
and a nontrivial zero-field limit follows if $\lim_{u\to\infty}\bar f(u)\sim u$, resulting in
\[
 \tau_{\text{AT}}(H=0)\sim\frac{1}{z}\:,
\]
in agreement with the previous regularization scheme.

\end{itemize}


\item[(ii)]{\bfseries The zero-temperature limit of the AT line for
$n=0$:}

Close to $T=0$ the loop-expansion is valid for $1/z\ll (T/T_c^{\text{mf}})^2\ll 1$, providing the result
\[
 \frac{H^2_{\text{AT}}}{(kT_c^{\text{mf}})^2}=
 \ln z+\ln\left(\frac{8}{9\pi}\,u\right)+O(u)
 \equiv \ln z+g(u)
 \qquad \text{with}\qquad u=\frac{1}{z}\,
 \left(\frac{T_c^{\text{mf}}}{T}\right)^2\:.
\]
The $T=0$ critical field is expected to be finite
for a system with finite connectivity $z$, 
in contrast to the SK model. This means that $\lim_{u
\to\infty}
g(u)$ must be finite, providing 
\[
 H^2_{\text{AT}}(T=0)= 
 (kT_c^{\text{mf}})^2\cdot [\ln z+g(\infty)]\:.
\]
\end{description}

\section{Perturbative RG for the cubic field theory}\label{RG}

MF theory 
and its perturbative corrections provide insight into the transition to the RSB phase. The renormalization group (RG) can also usually provide the correct phase diagrams and universal critical parameters for finite $d$, short-range systems ($z$ finite).
The $H=0$ case was initially studied by Wilson's RG, which identified a stable fixed point in the first order of the $\epsilon$-expansion,
$\epsilon=6-d$ \cite{Harris_Lubensky_Chen,Chen_Lubensky}. Later works extended the calculation of the critical exponents
$\eta$ and $\nu$ up to third order \cite{Gr85}.
An attempt of the RG study for the RS spin glass phase (again for $H=0$), with the result of finding its instability, was also done in \cite{PytteRudnick79}. Except in this last work,
 a single mass $m_1$ and cubic 
coupling $w_1$ were considered (see Eqs.~\eqref{LG} and \eqref{LI}), and it is the replicated paramagnet which becomes unstable on the
critical surface belonging to this stable fixed point.
This single critical mass is actually a direct consequence of the extra symmetry the replicated paramagnet has over the generic RS phase \cite{nucl}, thus resulting in the degeneracy of three different masses of the RS phase:
replicon, anomalous and longitudinal \cite{PytteRudnick79,BrMo79}.

The \emph{true} spin glass transition, i.e.\ the AT transition,
has a single critical mass, namely the replicon one $m_R$,
and only the two fully replicon cubic couplings $g_1$ and $g_2$
are different from zero (see Eq.~\eqref{cubic_with_g}). The first-order RG for this three-parameter model was worked out by Bray and Roberts \cite{BrRo} who found no stable fixed point when $d<6$. As for the case above $d=6$, the stable Gaussian fixed point has, somewhat unusually, a finite basin of attraction that vanishes as $d\to 6^+$ 
\cite{Moore_Bray_2011}. Although not specifically examined, this finite basin of attraction may exist in any high $d$, and physical systems outside of it may then not be attracted by the Gaussian fixed point.

These two parts of the parameter space---the replicated paramagnet and the fully replicon subspace---are closed under the RG iteration, and are both special cases of a more general RG system with three masses and eight cubic couplings (see
Eqs.~\eqref{LG},~\eqref{LI} and~\eqref{cubic_with_g}). The first-order RG equations in this large parameter space were presented for generic $n$ and for $n=0$ in Ref.\ \cite{Iveta}. The most important conclusion from this many-parameter RG is that there is
a critical AT surface in the crossover region around the zero-field fixed point over a range of dimensions $d\lesssim 6$, $d=6$, and $d\gtrsim 6$ \cite{PT,PRB_2017}.
\footnote{Note that all these contributions consider a pure cubic model, which is related to but not equal with the effective field theory proposed here.} The existence of the critical AT surface around the $H=0$ fixed point does \emph{not} 
contradict the lack of a stable AT-like fixed point: runaway trajectories for $g_1$ and $g_2$ are expected as the RG iterations push the system toward zero temperature.


\section{An unfinished story: transition to the RSB phase}

{\bfseries{Stable, strong coupling fixed points}:}
 Various lines of evidence support the existence of an AT-like transition in short range systems over a wide $d$ range. Examples include the numerical
work in $d=4$ \cite{Janus_2012_1}, Wilson's perturbative RG around the zero-field fixed point (Sec.\ \ref{RG}), and perturbative corrections
to MF theory (Sec.\ \ref{perturbative_part}).
Nevertheless, a theoretical understanding of the AT critical state is still lacking. The failure to find a stable nontrivial fixed point for $d<6$ in the one-loop perturbative
RG \cite{BrRo} and the runaway RG trajectories may be explained by a possible strong coupling fixed point (which is undetectable at one-loop level). Evidence for such a fixed point (stable over a range of $d$) has been found at two-loop level in Ref.\ \cite{Charbonneau_Yaida_2017}, supplemented by a
three-loop calculation and a resummation procedure~\cite{Charbonneau_2019}, but the situation remains inconclusive.

{\bfseries
Initial conditions for the RG iteration in $\mathbf{ d>6}$:}
For $d>6$ the Gaussian fixed point is stable, but its basin of attraction is finite \cite{Moore_Bray_2011}. This assessment refers to the fully replicon subspace, which is closed under the RG flow. Physical systems, however, when they are considered as initial conditions for an RG flow, usually lie
outside of this subspace. It is therefore nontrivial to predict the outcome of an RG iteration. In Fig.\ \ref{fig3} the AT line of the effective field theory, introduced in Sec.\ \ref{the_model},
is shown for some $d>6$ and $1/z\ll 1$; the perturbative study
in Sec.\ \ref{perturbative_part} is applied here. Three initial conditions on the AT line (where the exact replicon mass $\Gamma_R$ is zero) are considered: 
\begin{figure}
\centerline{\includegraphics[scale=0.4]{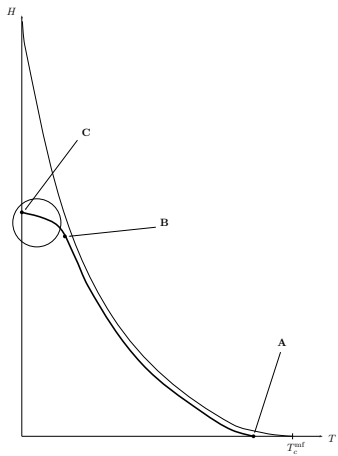}}
\caption{Phase diagram of the effective field theory in the
$H-T$ plane. Three states on the AT line are
considered as
initial conditions for the RG iteration (see text for details).
The encircled region is the nonperturbative part of the RS-to-RSB transition.
For comparison, the MF (tree-level) AT line is also displayed (narrow line).}
\label{fig3}
\end{figure} 

\begin{itemize}
 \item
 State A: $H=0$ endpoint of the AT line, which does not necessarily coincide with the critical point of the replicated paramagnet (see Sec.\ \ref{perturbative_part}). 
 This state (and those with $H\gtrsim 0$) is \emph{far} from the fully replicon subspace, because we have (by only showing the leading terms):
 \begin{gather*}
  g_1=1\,,\quad g_3=-1\,,\quad
  g_5=-1\,,\quad g_6=2\,,
  \quad \bar g_7=-\frac{3}{2}\,,\quad \bar g_8=\frac{1}{4}\,,\\[5pt]
  g_2\sim \frac{1}{z}\,,\qquad\bar g_4\sim \frac{1}{z}\,,\qquad
  m_R\sim -\frac{1}{z^2}\,,\qquad m_L\sim\frac{1}{z}\:;
 \end{gather*}
 see Ref.~\cite{Iveta} for the definitions of the \emph{bared} couplings which must be used when $n=0$. Since $g_i/\sqrt{z}\ll 1$ for all $i$,
 this state lies inside the perturbative region.
 From state A, RG iterations move the system toward the fully replicon subspace, driven by the hardening longitudinal mass ($m_L\to \infty$). Decoupling of the RG equations for $g_1$ and $g_2$ occurs when $m_L=\infty$. Although Ref.\ \cite{Moore_Read_2018} supposed that  the RG flow ends at the Gaussian fixed point $g_1=g_2=0$ when $d\gtrsim  6$, it is difficult to see this, and 
 a runaway flow to infinity is also conceivable.

 \item
 State B: The low-temperature
 end of the \emph{perturbative} AT line where $\frac{1}{z}
 \ll \left(\frac{T}{T_c^{\text{mf}}}\right)^2\ll 1$. To lighten the notation, let us define $\eta\equiv \left(\frac{T}{T_c^{\text{mf}}}\right)^2\ll1$. We have for the couplings:
 \[
  g_1=g_2=\frac{4}{5}\,\eta^{-1/2}\gg 1\qquad\text{whereas}\qquad 
  g_3,\:\bar g_4,\:g_5,\:g_6,\:\bar g_7,\:\bar g_8\sim\eta^{1/2}\ln \eta\ll1\:.
 \]
Because $g_i/\sqrt{z}\ll 1$ even for $i=1,\:2$, this initial state is still inside the perturbative regime. As for the masses, $m_L=O(1)$ and $m_R\sim- 1/\eta z\ll1$. Although this state is obviously dominated by the replicon mode, it is still somewhat outside the fully replicon subspace.
 
 \item State C: The encircled region in Fig.~\ref{fig3} shows the nonperturbative part of the AT line where $1/\eta z=O(1)$ (see Sec.\ \ref{perturbative_part}). As for the replicon couplings,
$g_1/\sqrt{z}=g_2/\sqrt{z}$ are also of order unity and we are out of the range where the perturbative RG is applicable.
State C is the zero-temperature limit of the AT line where
$1/\eta z\to \infty$. Considering these infinitely large
replicon couplings at $T=0$, one can certainly conclude that,
notwithstanding the correct phase diagram with the finite critical field at $T=0$,
the zero-temperature spin glass is not faithfully represented
by the effective field theory put forward here. One can speculate that regularization with the replica number $n\gtrsim 0$ may remedy the problem. Alternatively, 
the loop expansion around the Bethe lattice
(instead of the fully connected limit) at $T=0$ may provide a
solution to the problem \cite{loop_expansion_around_Bethe_lattice}.

 \end{itemize}
 
 \begin{acknowledgments}
  We are grateful to P. Charbonneau for his careful and valuable editorial work.
  One of us (T.T.) acknowledges financial support from the Hungarian Science Found (OTKA), No. K125171.
 \end{acknowledgments}

\bibliography{spinglass_for_RSB40}      
 
\end{document}